\documentstyle[epsfig,mathptm]{mn}
\newif\ifAMStwofonts
\AMStwofontstrue                
\DeclareMathVersion{bold}                    
\DeclareMathAlphabet{\mathbfit}{OT1}{cmr}{bx}{it}
\SetMathAlphabet\mathbfit{bold}{OT1}{cmr}{bx}{it}
\DeclareMathAlphabet{\mathbfss}{OT1}{cmss}{bx}{n}
\SetMathAlphabet\mathbfss{bold}{OT1}{cmss}{bx}{n}
\ifAMStwofonts
  \ifCUPmtlplainloaded \else
    \DeclareSymbolFont{UPM}{U}{eur}{m}{n}
    \SetSymbolFont{UPM}{bold}{U}{eur}{b}{n}
    \DeclareSymbolFont{AMSa}{U}{msa}{m}{n}
    \DeclareMathSymbol{\upi}{0}{UPM}{"19}
    \DeclareMathSymbol{\umu}{0}{UPM}{"16}
    \DeclareMathSymbol{\upartial}{0}{UPM}{"40}
    \DeclareMathSymbol{\leqslant}{3}{AMSa}{"36}
    \DeclareMathSymbol{\geqslant}{3}{AMSa}{"3E}

     \let\le=\leqslant
     \let\ge=\geqslant
  \fi
\fi
\def\mat#1{{\mathbfss #1}}
\def\pd#1#2{{\upartial #1 \over \upartial #2}}
\def\spd#1#2#3{{\upartial ^2 #1 \over \upartial #2 \upartial #3}}
\title[The entropic prior for positive/negative distributions]
{The entropic prior for distributions with positive and negative
values}
\author[M.P.~Hobson and A.N.~Lasenby]
{M.P.~Hobson and A.N.~Lasenby\\
Mullard Radio Astronomy Observatory,
Cavendish Laboratory, Madingley Road, Cambridge, CB3 0HE}
\date{Accepted 1998 April 1. Received 1998 March 30; in original form 
1997 December 22}
\begin{document}
\maketitle
\begin{abstract}
The maximum entropy method has been used to reconstruct images in a
wide range of astronomical fields, but in its traditional form it is
restricted to the reconstruction of strictly positive distributions.
We present an extension of the standard method to include
distributions that can take both positive and negative values. The
method may therefore be applied to a much wider range of astronomical
reconstruction problems.  In particular, we derive the form of the
entropy for positive/negative distributions and use direct counting
arguments to find the form of the entropic prior. We also derive the
measure on the space of positive/negative distributions, which allows
the definition of probability integrals and hence the proper
quantification of errors. 
\end{abstract}


\section{Introduction}
\label{intro}

The maximum entropy method (MEM) has been applied to numerous image
reconstruction problems in astronomy and various other fields (for
reviews see Gull \& Skilling (1984), Titterington (1985) and Narayan
\& Nityananda (1986)).  In its standard form, however, it is only
applicable to distributions that are stricly positive.  For most
problems in astronomy, one is concerned with the reconstruction of the
intensity of an image from some convolved and noisy data, and so the
restriction of positivity is not a problem.  Nevertheless, more
recently, the astronomical community has become interested in the use
of Bayesian methods to reconstruct images of the temperature
fluctuations in the cosmic microwave background.  In order to avoid
the introduction of any bias into the reconstruction of such an image,
it is desirable to reconstruct the temperature fluctuations about some
mean level, rather than introducing some arbitrary offset to ensure
the positivity. Clearly such a problem requires the reconstruction
to take both positive and negative values about the mean level, and
this has lead to the consideration of how to apply MEM
to such problems (Maisinger, Hobson \& Lasenby 1997; Hobson et
al. 1997; Jones et al. 1997).

The theoretical foundations of MEM, when applied to strictly
positive distributions, are discussed in some detail by Skilling
(1988, 1989).  From very general notions of subset independence,
coordinate independence, system independence and scaling, Skilling (1988)
shows that the assignment of a positive distribution $f(x)$, given
some definitive but incomplete constraints on it, must be made by
maximising over $f$ (subject to the imposed constraints) the entropy
functional $S[f,m]$ given by
\begin{equation}
S[f,m]=\int f(x)-m(x)-f(x)\ln\left[\frac{f(x)}{m(x)}\right] \,{\rm d}x.
\label{mem1}
\end{equation}
In this expression $m(x)$ is the Lebesgue measure associated with $x$,
which must be given before the integral can be defined. In practice,
however, $m(x)$ is more usually interpreted as the {\em model} to
which $f(x)$ defaults in the absence of any constraints. Indeed it is
easily shown that $S[f,m]$ possesses a single global maximum at
$f(x)=m(x)$. It is also common in practice that the distribution $f(x)$ is
digitised on to $N$ equivalent cells to form the vector
$\mat{f}=(f_1,f_2,\ldots,f_N)$ and a similar digitisation is also
performed for the model $m(x)$.  In this case, the integral in
(\ref{mem1}) is simply replaced by a sum over the cells, so that
\begin{equation}
S[\mat{f},\mat{m}]
=\sum_{i=1}^N \left[f_i-m_i-f_i\ln\left(\frac{f_i}{m_i}\right)\right].
\label{mem2}
\end{equation}

The arguments leading to (\ref{mem2}) do not, however, address the
reliability of this assignment of the distribution $\mat{f}$. In other
words, we are not able to quantify how much better is this assignment
of $\mat{f}$ as compared to any other distribution permitted by
the imposed constraints.
Moreover, experimental data are usually noisy
and so do not constitute testable constraints on the distribution
$\mat{f}$, but instead define a likelihood function
$\Pr(\mbox{data}|\mat{f})$. Thus, in order to apply a proper Bayesian
analysis, it is necessary to derive the full entropic prior
probability distribution $\Pr(\mat{f}|\mat{m})$.  Following Skilling
(1989), this leads us to consider probability integrals over domains
$V$ in the space of distributions. Assuming $\mat{f}$ to be digitised on
to $N$ cells, as described above, we write
\[
\Pr(\mat{f} \in V|\mat{m})
=\int_V {\rm d}^N\mat{f}\, M(\mat{f}) \Pr(\mat{f}|\mat{m}),
\]
where $M(\mat{f})$ is the measure on the space of positive
distributions. In order that we identify the most probable
distribution with that assigned by maximising the entropy, we must
have
\begin{equation}
\Pr(\mat{f} \in V|\mat{m})
=\frac{1}{Z_S(\alpha,\mat{m})}
\int_V {\rm d}^N\mat{f}\,M(\mat{f})\,\Phi(\alpha S[\mat{f},\mat{m}]),
\label{genprior}
\end{equation}
where $\Phi$ is a monotonic function of the dimensionless argument
$\alpha S[\mat{f},\mat{m}]$. Since $S[\mat{f},\mat{m}]$ has dimensions
of integrated $\mat{f}$, then $\alpha$ is also dimensional and is not an
absolute constant. The factor $Z_S(\alpha,\mat{m})$ is required to
ensure that $\Pr(\mat{f}|\mat{m})$ is properly normalised and is given
by
\[
Z_S(\alpha,\mat{m})=
\int_\infty {\rm d}^N\mat{f}\, M(\mat{f}) \,\Phi(\alpha S[\mat{f},\mat{m}]).
\]

In order to find the function $\Phi$ it is necessary to consider a
specific distribution $\mat{f}$ that obeys the general notions of
subset independence etc. mentioned above, but for which the integral
$\Pr(\mat{f} \in V|\mat{m})$ can be calculated directly. The
traditional example used (Frieden 1972, Gull \& Daniell 1979, Jaynes
1986) is
that of a distribution $\mat{f}$ created by a `team of monkeys' throwing
balls randomly at the cells $i=1,2,\ldots,N$. As shown by Skilling
(1989), this leads to the identifications
\begin{eqnarray*}
\Phi(\alpha S[\mat{f},\mat{m}]) & = & \exp(\alpha S[\mat{f},\mat{m}]),\\
Z_S(\alpha,\mat{m}) & = & (2\pi/\alpha)^{N/2},
\end{eqnarray*}
as well as yielding an expression for the measure on the space of
positive distributions which is given by
\[
M(\mat{f})=\prod_{i=1}^N f_i^{-1/2},
\]
A knowledge of the measure is important, since it
enables us to define probability integrals over domains in the
space of distributions.

In this paper, we discuss the extension of MEM to distributions that
can take both positive and negative values.  The resulting formalism
is clearly applicable to the reconstruction of temperature
fluctuations in the microwave background, as discussed above, but may
also be applied to a much wider range of image reconstruction
problems. In particular, we present derivations of the general form of
the entropy for positive/negative distributions and calculate the
corresponding entropic prior probability distribution.

\section{The entropy of positive/negative distributions}
\label{pnentropy}

It is clear from the logarithmic term in the entropy functional
(\ref{mem2}) that negative values are not permitted in the (digitised)
distribution $\mat{f}$. Nevertheless, the problem of assigning
distributions that can take both positive and negative values has been
briefly considered by Gull \& Skilling (1990) for certain
special cases. In this section, we extend their method to find the
form of the entropy for general positive/negative distributions.

The central idea is to express
the general (positive/negative) distribution $\mat{h}$ as the
difference of two strictly positive distributions $\mat{f}$ and
$\mat{g}$, i.e.
\[
\mat{h}=\mat{f}-\mat{g}.
\]
If $\mat{m}_f$ and $\mat{m}_g$ are separate models for $\mat{f}$
and $\mat{g}$ respectively, then the total entropy can be written
as
\begin{eqnarray}
S[\mat{h},\mat{m}_f,\mat{m}_g]
& = & 
\sum_{i=1}^N \left\{f_i-(m_f)_i-f_i\ln\left[\frac{f_i}{(m_f)_i}\right]\right\}
\nonumber \\
& + & 
\sum_{i=1}^N \left\{g_i-(m_g)_i-g_i\ln\left[\frac{g_i}{(m_g)_i}\right]\right\}.
\label{pnent1}
\end{eqnarray}
Our aim is to derive a form for $S[\mat{h},\mat{m}_f,\mat{m}_g]$ in
which only $\mat{h}$ appears, rather than the two positive
distributions $\mat{f}$ and $\mat{g}$. We cannot, however, hope to
replace the models $\mat{m}_f$ and $\mat{m}_g$ by a single
positive/negative model $\mat{m}_h$ (say), since such a distribution
could no longer be considered as an integration measure. Nevertheless,
we can still consider the difference $\mat{m}_f-\mat{m}_g$ as the
model for the distribution $\mat{h}$.

We may remove the explicit dependence of
$S[\mat{h},\mat{m}_f,\mat{m}_g]$ on the
positive distributions $\mat{f}$ and $\mat{g}$ by
applying continuity constraints on the entropy functional.  Since
$\mat{h}=\mat{f}-\mat{g}$, we have
\begin{eqnarray*}
\pd{S}{f_i} & = & \sum_{k=1}^N \pd{S}{h_k}\pd{h_k}{f_i} = \pd{S}{h_i} \\
\pd{S}{g_i} & = & \sum_{k=1}^N \pd{S}{h_k}\pd{h_k}{g_i} = -\pd{S}{h_i},
\end{eqnarray*}
from which we obtain the constraint
\begin{equation}
\pd{S}{f_i}+\pd{S}{g_i}=0.
\label{const1}
\end{equation}
It is straightforward to show that
\[
\pd{S}{f_i} = -\ln\left[\frac{f_i}{(m_f)_i}\right],
\]
and a similiar expression exists for $\upartial S/\upartial g_i$. 
Hence (\ref{const1}) becomes
\[
\ln\left[\frac{f_i}{(m_f)_i}\right]+\ln\left[\frac{g_i}{(m_g)_i}\right]=0,
\]
and thus we obtain the  constraint $f_ig_i=(m_f)_i(m_g)_i$. Using
this constraint to substitute separately for $f_i$ and
$g_i$ in the relation $h_i=f_i-g_i$,
and remembering that $f_i$, $g_i$, $(m_f)_i$ and $(m_g)_i$ are all 
required to be strictly positive, we obtain
\begin{eqnarray}
f_i & = & {\textstyle\frac{1}{2}}(\psi_i + h_i), \label{fdef}\\
g_i & = & {\textstyle\frac{1}{2}}(\psi_i - h_i). \label{gdef}
\end{eqnarray}
where $\psi_i=[h_i^2 + 4(m_f)_i(m_g)_i]^{1/2}$.
Finally, substituting the relationships (\ref{fdef}) and (\ref{gdef})
into the expression 
for the entropy given in (\ref{pnent1}), after a little algebra 
we find the form of the entropy for the positive/negative
distribution $\mat{h}$ is given by
\begin{equation}
S[\mat{h},\mat{m}_f,\mat{m}_g] = 
\sum_{i=1}^N \left\{\psi_i-(m_f)_i-(m_g)_i
-h_i\ln\left[\frac{\psi_i+h_i}{2(m_f)_i}\right]\right\}.
\label{pnentfinal}
\end{equation}

\section{The entropic prior for positive/negative distributions}
\label{pnprior}

As discussed in Section \ref{intro}, its still remains to quantify the
assignment of the positive/negative distribution $\mat{h}$, by
deriving the entropic prior probability
$\Pr(\mat{h}|\mat{m}_f,\mat{m}_g)$. We may calculate this prior by
again considering the specific case in which $\mat{h}$ is created using
the traditional team of monkeys, but now throwing {\em two} types of
`balls', for example red ones and blue ones (each of quantum size $q$)
randomly into the cells, with Poisson expectations $\lambda_i$ and
$\mu_i$ respectively. The distribution $\mat{h}$ is then found by
considering the number of red balls $r_i$ minus the number of blue
balls $b_i$ in each cell. In particular, we define $f_i=r_iq$ and
$g_i=b_iq$, so that $h_i=n_iq$ where $n_i=r_i-b_i$. We also define the
models for the positive and negative portions of the distribution
$\mat{h}$ in a similar way as $(m_f)_i=\lambda_iq$ and
$(m_g)_i=\mu_iq$.

We may now consider the space of $\mat{h}$ to be constructed from
microcells of volume $q^N$, each one associated with one lattice point
of integers $\mat{n}=(n_1,n_2,\ldots,n_N)$, where each integer may be
positive or negative (or zero). Since each cell is independent of the
others, the probability of obtaining a given set of occupation
numbers is given by the product of the probability distributions for
the occupation number in each cell separately, i.e.
\[
\Pr(\mat{n}|\blambda,\bmu)
=\prod_{i=1}^N \Pr(n_i|\lambda_i,\mu_i).
\]

We see that in order to proceed further, we must consider the
probability distribution of the difference of two independent Poisson
variates. This distribution is derived in the Appendix and is
\[
\Pr(n_i|\lambda_i,\mu_i)=
\exp[-(\lambda_i+\mu_i)]
\left(\frac{\lambda_i}{\mu_i}\right)^{n_i/2} I_{n_i}(2\sqrt{\lambda_i\mu_i}),
\]
where $I_n(x)$ is the modified Bessel function of order $n$.
Thus, as we let the quantum size $q \to 0$, the probability that the
distribution $\mat{h}$ lies in some domain $V$ of distribution
space is given by
\[
\Pr(\mat{h} \in V|\mat{m}_f,\mat{m}_g) =
\]
\begin{equation}
\hspace{1cm}
\int_V \frac{{\rm d}^N\mat{h}}{q^N} 
\prod_{i=1}^N
\exp[-(\lambda_i+\mu_i)]
\left(\frac{\lambda_i}{\mu_i}\right)^{n_i/2} I_{n_i}(2\sqrt{\lambda_i\mu_i}).
\label{prionh}
\end{equation}

In the limit $q \to 0$, the integers $n_i$ will in general become
very large, and so we are led to consider the uniform asymptotic
expansion of a modified Bessel functions of large order, which reads
(Abramowitz \& Stegun 1972)
\begin{equation}
I_n(nx) \sim \frac{1}{(2\pi n)^{1/2}}\frac{\exp(n\beta)}{(1+x^2)^{1/4}},
\label{bessiexp}
\end{equation}
where the variable $\beta$ is given by
\begin{equation}
\beta = (1+x^2)^{1/2} + \ln\left[\frac{x}{1+(1+x^2)^{1/2}}\right].
\label{beta}
\end{equation}
Setting $x=2\sqrt{\lambda\mu}/n$ in (\ref{bessiexp}) and (\ref{beta})
and rearranging, we obtain
\[
I_n(2\sqrt{\lambda\mu}) \sim
\frac{1}{(2\pi\phi)^{1/2}}
\left[\frac{2(\lambda\mu)^{1/2}}{n+\phi}\right]^n\exp(\phi).
\]
where we have defined the quantity
$\phi=(n^2+4\lambda\mu)^{1/2}$.
Substituting this result into (\ref{prionh}), we find
\begin{equation}
\Pr(\mat{h} \in V|\mat{m}_f,\mat{m}_g)
= \int_V \frac{{\rm d}^N\mat{h}}{\prod_{i=1}^N (2\pi q^2\phi_i)^{1/2}}
\exp\left[\xi(\bphi,\blambda,\bmu)\right],
\label{tempres1}
\end{equation}
where the function  $\xi(\bphi,\blambda,\bmu)$ is given by
\begin{eqnarray*}
\xi(\bphi,\blambda,\bmu)
\!\!\!& = &\!\!\!
\exp\left[\sum_{i=1}^N (\phi_i-\lambda_i-\mu_i)\right] 
\left(\frac{\lambda_i}{\mu_i}\right)^{n_i/2} 
\left[\frac{2(\lambda_i\mu_i)^{1/2}}{n_i+\phi_i}\right]^{n_i} \\
\!\!\!& = &\!\!\!
\exp\left[\sum_{i=1}^N (\phi_i-\lambda_i-\mu_i)\right] 
\exp\left[-\sum_{i=1}^N
n_i\ln\left(\frac{\phi_i+n_i}{2\lambda_i}\right)
\right]. 
\end{eqnarray*}
Collecting together the exponential factors and writing the resulting
expression in terms of $\mat{h}$, $\mat{m}_f$ and
$\mat{m}_g$, we obtain
\begin{equation}
\xi(\bphi,\blambda,\bmu) =
\frac{1}{q}\sum_{i=1}^N \left\{\psi_i-(m_f)_i-(m_g)_i
-h_i\ln\left[\frac{\psi_i+h_i}{2(m_f)_i}\right]\right\},
\label{xidef}
\end{equation}
where $\psi_i=\phi_i q = [h_i^2 + 4(m_f)_i(m_g)_i]^{1/2}$, and on the
RHS we recognise the entropy functional given in
(\ref{pnentfinal}). Thus, substituting (\ref{xidef}) into
(\ref{tempres1}), we find
\begin{equation}
\Pr(\mat{h} \in V|\mat{m}_f,\mat{m}_g)
=\int_V \frac{{\rm d}^N\mat{h}}{\prod_{i=1}^N (2\pi\psi_i)^{1/2}}
\frac{\exp(S(\mat{h},\mat{m}_f,\mat{m}_g)/q)}
{(2\pi q)^{N/2}}.
\label{entprior2}
\end{equation}

We may now compare this result with our earlier expression (\ref{genprior}).
Aside from multiplicative constants which can be defined to be unity,
we therefore identify the quantum size $q$ of the `balls' as
$q=1/\alpha$. As in the case of a strictly positive distribution, the 
monotonic function $\Phi$ of the entropy functional
is identified as 
\[
\Phi(\alpha S(\mat{h},\mat{m}_f,\mat{m}_g))
=\exp(\alpha S(\mat{h},\mat{m}_f,\mat{m}_g)),
\]
which together with the normalisation constant
\[
Z_S(\alpha,\mat{m}_f,\mat{m}_g)=(2\pi/\alpha)^{N/2},
\]
defines the entropic prior on the space of positive/negative
distributions. Furthermore, from (\ref{entprior2}), we also identify
the measure on the space of positive/negative distributions as
\[
M(\mat{h}) = \prod_{i=1}^{N} \psi_i^{-1/2},
\]
which enables us to define probability integrals over domains $V$
in this space.

As discussed by Skilling (1989), a natural interpretation of the
measure is as the invariant volume $({\rm det}~\mat{G})^{1/2}$
of a metric $\mat{G}$ defined on the space. Thus the natural
metric for the space of positive/negative distributions is simply
\[
\mat{G}={\rm diag}(1/\psi_1,1/\psi_2,\ldots,1/\psi_N).
\]
Furthermore, it is 
straightforward to show from (\ref{pnentfinal}) that the elements of
the Hessian matrix of the entropy functional are given by
\[
\spd{S}{h_i}{h_j} = -\frac{1}{\psi_i}\delta_{ij}
\]
so that, as in the case of positive-only distributions, we may write
the metric as
\[
\mat{G} =-\nabla_\mat{h}\nabla_{\mat h}S(\mat{h},\mat{m}_f,\mat{m}_g).
\]
Thus the metric on the space of positive/negative distributions at any
given point is equal to minus the Hessian matrix of the entropy
functional at that point.

\section{Conclusions}
\label{conc}

We have presented derivations of the form of the entropy for
distributions that can take both positive and negative values. We also
use direct counting arguments to calculate the form of the entropic
prior for such distributions. This enables the maximum entropy method
to be applied directly to the reconstruction of general
positive/negative distributions. This method will therefore be of use in
a much wider range of astronomical image reconstruction problems.
The calculations presented here also
yield the form of the measure on the space of positive/negative
distributions, which enables the calculation of probability integrals
over domains in this space, and therefore the proper quantification of
errors on the reconstructed distribution.

\subsection*{ACKNOWLEDGMENTS}
We thank Steve Gull and John Skilling for some useful
discussions.

\appendix

\section{The difference of two independent Poisson variates}

Let us consider two {\em independent} random variables $X$ and $Y$.
If $X$ is Poisson distributed with mean $\lambda$ and $Y$ is Poisson
distributed with mean $\mu$ then
\begin{eqnarray}
\Pr(X=n) & = & \frac{\lambda^n}{n!}\exp(-\lambda), \label{poissx}\\
\Pr(Y=m) & = & \frac{\mu^m}{m!}\exp(-\mu).\label{poissy}
\end{eqnarray}

If we now consider the variable $Z=X-Y$, then its probability
distribution is given by
\begin{equation}
\Pr(Z=n)=
\cases{
$${\displaystyle \sum_{r=0}^{\infty} \Pr(X=n+r)\Pr(Y=r)}$$  
& for $n \ge 0$ \cr
$$\stackrel{~}
{{\displaystyle \sum_{r=0}^{\infty} \Pr(X=r)\Pr(Y=|n|+r)}}$$ 
& for $n \le 0$. \cr
}
\label{probz}
\end{equation}
Substituting the expressions (\ref{poissx}) and (\ref{poissy}) into
(\ref{probz}), we obtain
\begin{equation}
\Pr(Z=n)=
\cases{
$${\displaystyle \exp[-(\lambda+\mu)]
\sum_{r=0}^{\infty} \frac{\lambda^{n+r}\mu^r}{(n+r)!r!}}$$
& for $n \ge 0$ \cr
$$\stackrel{~}
{{\displaystyle \exp[-(\lambda+\mu)]
\sum_{r=0}^{\infty} \frac{\lambda^r\mu^{|n|+r}}{r!(|n|+r)!}}}$$
& for $n \le 0$.\cr}
\label{probz2}
\end{equation}
Provided $n$ is a non-negative integer, however, the modified Bessel
function of order $n$ is given by
\[
I_n(x)={\rm i}^{-n}J_n({\rm i}x)
=\sum_{r=0}^\infty \frac{(x/2)^{n+2r}}{r!(n+r)!},
\]
and so we may write the first sum in (\ref{probz2}) as
\begin{eqnarray*}
\lambda^n \sum_{r=0}^{\infty} \frac{(\lambda\mu)^r}{(n+r)!r!} 
& = & \lambda^n I_n(2\sqrt{\lambda\mu})(\lambda\mu)^{-n/2} \\
& = & \left(\frac{\lambda}{\mu}\right)^{n/2} I_n(2\sqrt{\lambda\mu}).
\end{eqnarray*}
Writing the second sum in (\ref{probz2}) in a similar way, we thus obtain
\[
\Pr(Z=n)=
\cases{
$${\displaystyle \exp[-(\lambda+\mu)]
\left(\frac{\lambda}{\mu}\right)^{n/2} I_n(2\sqrt{\lambda\mu})}$$
& for $n \ge 0$ \cr
$$\stackrel{~}
{{\displaystyle \exp[-(\lambda+\mu)]
\left(\frac{\mu}{\lambda}\right)^{|n|/2} I_{|n|}(2\sqrt{\lambda\mu})}}$$
& for $n \le 0$.}
\]
If $n$ is an integer, however, $I_{-n}(x)=I_n(x)$ and so the final
expression for the probability distribution of $Z$ is given for all
integers $n$ by

\begin{equation}
\Pr(Z=n)=
\exp[-(\lambda+\mu)]
\left(\frac{\lambda}{\mu}\right)^{n/2} I_n(2\sqrt{\lambda\mu}).
\label{probzfinal}
\end{equation}
The probability distribution for 
difference of two independent Poisson variates has also been
considered by Stuart \& Ord (1994), but only for the special
case $\lambda=\mu$ and with the result expressed only as an infinite
series.

A useful check that (\ref{probzfinal}) is correctly normalised is
provided by considering the generating function for modified Bessel
functions, which reads 
\[
\sum_{n=-\infty}^\infty I_n(x)t^n
=\exp\left[{\textstyle\frac{1}{2}}x\left(t+\frac{1}{t}\right)\right].
\]
Substituting $t=\sqrt{\lambda/\mu}$ and $x=2\sqrt{\lambda\mu}$ into
this expression, we find
\begin{eqnarray*}
\sum_{n=-\infty}^\infty I_n(2\sqrt{\lambda\mu})
\left(\frac{\lambda}{\mu}\right)^{n/2}
& = & 
\exp\left[\sqrt{\lambda\mu}
\left(\sqrt{\frac{\lambda}{\mu}}+\sqrt{\frac{\mu}{\lambda}}\right)\right] \\
& = & \exp(\lambda+\mu).
\end{eqnarray*}
Hence, substituting this result into (\ref{probzfinal}), we find that
as required
\[
\sum_{n=-\infty}^{\infty} \Pr(Z=n) =
\exp[-(\lambda+\mu)]\exp(\lambda+\mu) = 1.
\]

\bsp 
\label{lastpage}

\begin{thebibliography}{}

\bibitem[\protect\citename{Abramowitz \& Stegun }1972]{abramowitz72}
Abramowitz M., Stegun I.A., 1972, Handbook of Mathematical Functions.
Dover, New York.
\bibitem[\protect\citename{Frieden }1972]{frieden72}
Frieden B.R., 1972, J.~Opt.~Soc.~Am., 62, 511
\bibitem[\protect\citename{Gull \& Daniell }1979]{gull79}
Gull S.F., Daniell G.J., 1979, in van Schooneveld C., ed., Image
Formation from Coherence Functions in Astronomy. Riedel, p.~219
\bibitem[\protect\citename{Gull \& Skilling }1984]{gull84}
Gull S.F., Skilling J., 1984, IEE Proc., 131(F), 646
\bibitem[\protect\citename{Gull \& Skilling }1990]{gull90}
Gull S.F., Skilling J., 1990, The MEMSYS5 Users' Manual. Maximum
Entropy Data Consultants Ltd, Royston.
\bibitem[\protect\citename{Hobson et al. }1997]{hobson97}
Hobson M.P., Jones A.W., Lasenby A.N., Bouchet F.R., 1997, MNRAS, submitted
\bibitem[\protect\citename{Jaynes }1986]{jaynes86}
Jaynes E.T., 1986, in Justice J.H., ed., Maximum Entropy and Bayesian
Methods in Applied Statistics. Cambridge University Press, Cambridge, p.~26
\bibitem[\protect\citename{Jones et al. }1997]{aled}
Jones A.W., Hancock S., Lasenby A.N., Davies R.D., Guti\'{e}rrez C.M.,
Rocha G., Watson R.A., Rebolo R., 1997, MNRAS, submitted
\bibitem[\protect\citename{Maisinger et al. }1997]{maisinger97}
Maisinger K., Hobson M.P., Lasenby A.N., 1997, MNRAS, 290, 313
\bibitem[\protect\citename{Narayan \& Nityananda }1986]{narayan86}
Narayan R., Nityananda R., 1986, ARA\&A, 24, 127
\bibitem[\protect\citename{Skilling }1988]{skilling88}
Skilling J., 1988, in Erickson G.J., Smith C.R., eds, Maximum Entropy 
and Bayesian Methods in Science and Engineering. Kluwer, Dordrecht, p.~173
\bibitem[\protect\citename{Skilling }1989]{skilling89}
Skilling J., 1989, in Skilling J., ed., Maximum Entropy and Bayesian
Methods. Kluwer, Dordrecht, p.~53
\bibitem[\protect\citename{Stuart \& Ord }1994]{stuart94}
Stuart A., Ord K.J., 1994, Kendall's Advanced Theory of Statistics
(Volume I). Edward Arnold, London
\bibitem[\protect\citename{Titterington }1985]{titterington85}
Titterington D.M., 1985, A\&A, 144, 381


\end{thebibliography}
\end{document}